\documentclass[fleqn,10pt]{wlscirep}
\usepackage[utf8]{inputenc}
\usepackage[T1]{fontenc}
\usepackage{upgreek}

\title{Exciton-to-trion conversion as a control mechanism for valley polarization in room-temperature monolayer WS$_\text{2}$}

\author[1,+]{Joris J. Carmiggelt}
\author[1,+]{Michael Borst}
\author[1,*]{Toeno van der Sar}
\affil[1]{Department of Quantum Nanoscience, Kavli Institute of Nanoscience, Delft University of Technology, Lorentzweg 1, 2628 CJ Delft, The Netherlands}

\affil[*]{t.vandersar@tudelft.nl}

\affil[+]{These authors contributed equally to this work}


\begin{abstract}
Transition metal dichalcogenide (TMD) monolayers are two-dimensional semiconductors with two valleys in their band structure that can be selectively addressed using circularly polarized light. Their photoluminescence spectrum is characterized by neutral and charged excitons (trions) that form a chemical equilibrium governed by the net charge density. Here, we use chemical doping to drive the conversion of excitons into trions in $\text{WS}_{2}$ monolayers at room temperature, and study the resulting valley polarization via photoluminescence measurements under valley-selective optical excitation. We show that the doping causes the emission to become dominated by trions with a strong valley polarization associated with rapid non-radiative recombination. Simultaneously, the doping results in strongly quenched but highly valley-polarized exciton emission due to the enhanced conversion into trions. A rate equation model explains the observed valley polarization in terms of the doping-controlled exciton-trion equilibrium. Our results shed light on the important role of exciton-trion conversion on valley polarization in monolayer TMDs.
\end{abstract}
\begin{document}

\flushbottom
\maketitle
%
%
\thispagestyle{empty}

\section*{Introduction}

Transition metal dichalcogenide (TMD) monolayers are direct-bandgap semiconductors of which the conduction and valence band extrema consist of two valleys \cite{Mak2010,Splendiani2010}. The broken inversion symmetry of the lattice gives rise to optical selection rules that enable valley-selective, inter-band excitation of electrons using circularly polarized light \cite{Mak2012,Cao2012,Zeng2012}. A strong Coulomb interaction results in the subsequent formation of excitons\cite{Mueller2018}, which maintain a valley polarization that is determined by the ratio between the intervalley scattering time and the exciton lifetime\cite{Mak2012,McCreary2017}. Such valley-polarized excitons have been proposed as carriers of information and play a central role in the field of valleytronics \cite{Mak2018,Schaibley2016}. As such, understanding the processes that govern the exciton lifetime and associated valley polarization is important for assessing the potential applicability of valley-polarized excitons in devices.

Under optical excitation, a charge-density-controlled chemical equilibrium between neutral and charged excitons (trions) forms in a TMD monolayer \cite{Lien2019,Ross2013,Siviniant1999}. The conversion into trions reduces the exciton lifetime\cite{HL2020} and may therefore be expected to lead to a large valley polarization of excitons that are created via valley-selective optical pumping, but demonstrating this effect has thus far remained elusive. 

The charge density of TMD monolayers can be controlled via electrostatic gating or chemical doping \cite{Lien2019, Mak2013,Ross2013,Tanoh2019,Wang2019,Amani2015,Peimyoo2014,Mouri2013,Tongay2013}. While electrostatic gating is a flexible technique that allows a continuous change of the charge density \cite{Mak2013,Ross2013, Lien2019}, chemical doping provides a convenient alternative that requires no microfabrication and is well suited for achieving high doping levels \cite{Tanoh2019,Wang2019,Amani2015,Peimyoo2014,Mouri2013,Tongay2013}. Here, we study the valley polarization of excitons and trions in monolayer $\text{WS}_{2}$ and show that chemical doping via aromatic anisole (methoxy-benzene) quenches the exciton photoluminescence and causes the spectrum to become dominated by trions with a strong valley polarization. A spatial study of the remaining exciton emission shows that also the excitons attain a strong valley polarization, which we attribute to the rapid doping-induced conversion into trions. We extend a rate equation model describing exciton-trion conversion \cite{Lien2019} to include the two valleys and use it to explain the observed valley polarization in terms of the doping-controlled chemical equilibrium between excitons and trions. 

\section*{Results}

When doping a TMD monolayer using aromatic molecules such as anisole, Hard Soft Acid Base (HSAB) theory allows predicting whether the dopant will be n- or p-type \cite{Wang2019}. Electrons hop between the adsorbed molecules ($A$) and the monolayer ($B$) to compensate for the difference in chemical potential $\mu$ between both systems\cite{Pearson1992}. The chemical hardness $\eta$ of the materials determines how quickly an equilibrium is reached, leading to an average number of transferred electrons per molecule $\Delta N$:
\begin{equation}
\Delta N=\frac{\mu_A-\mu_B}{\eta_A+\eta_B}.
\end{equation}
For both anisole and monolayer $\text{WS}_{2}$, the chemical potential and chemical hardness has been calculated using density functional theory\cite{Camacho-mendoza2018,Zhuang2013}. Using these values (Supplementary Section S1) we find $\Delta N=0.22$, such that we expect the monolayer to be n-doped upon physisorption of anisole molecules (Figure 1a).

To study the effect of chemical doping with anisole on the valley polarization properties of $\text{WS}_{2}$, we start by characterizing the photoluminescence of exfoliated $\text{WS}_{2}$ monolayers on 280 nm Si/$\textrm{SiO}_{2}$ substrates. The emission spectrum of an as-prepared monolayer shows the characteristic bright exciton resonance at 2.01 eV (Figure 1b, black line)\cite{McCreary2016}. After chemical doping by a two-hour treatment in liquid anisole at 70 C$^\circ$, the bright exciton resonance is strongly quenched and only a weak emission peak that is red-shifted by $\Delta E=23$ meV remains (Figure 1b, red line). Because the increased binding energy of trions compared to excitons should lead to such a red shift\cite{Mak2013} and the expected n-type doping by the anisole molecules should favour trion formation, we attribute this peak to emission associated with trions. This conclusion is further supported by spatial studies of emission spectra showing both exciton and trion components that we will describe below. As expected, the trion emission is weak due to its long radiative lifetime and strong non-radiative decay attributed to Auger recombination\cite{Lien2019,Hanbicki2016,Kurzmann2016}.

Doping by adsorbed carbon-hydrogen groups\cite{Zhang2019} was previously shown to result in an increase of the longitudinal acoustic LA(M) mode in the Raman spectrum of WS$_2$ monolayers. Our treatment causes a similar increase of the LA(M) Raman mode (Figure 1c), which we therefore attribute to the adsorption of anisole molecules. In addition, we find that the double-resonance 2LA(M) mode remains unaffected by the doping, indicating that our treatment does not significantly change the monolayer's electronic structure\cite{Berkdemir2013}. 

To study the valley polarization of chemically-doped WS$_2$ monolayers, we use near-resonant excitation with a 594 nm circularly polarized, continuous-wave laser that is focussed to a diffraction-limited spot. The resulting photoluminescence is polarization filtered and collected using a home-built confocal microscope (see Methods). Before detecting the emission with an avalanche photodiode (APD), we apply a spectral bandpass filter with a transmission window centred around the exciton and trion resonances (see the shaded area in Figure 1b).

We quantify the valley polarization $\rho$ via polarization-resolved photoluminescence measurements according to
\begin{equation}
\rho=\frac{I_{\sigma_+}-I_{\sigma_-}}{I_{\sigma_+}+I_{\sigma_+}}.
\end{equation}
Here, $I_{\sigma_+}$ and $I_{\sigma_-}$ represent the intensities of the right- and left-handed emission by the sample under $\sigma_+$ excitation and the total photoluminescence is given by $I=I_{\sigma_+}+I_{\sigma_-}$. By scanning the sample while detecting its emission using the APD, we make photoluminescence and valley-polarization maps of our flakes, before and after treating them.

Before the anisole treatment, the photoluminescence is characterized by bright exciton emission (Figure 2a, left panel) with no valley polarization (Figure 2b, left panel). Strikingly, the trion emission that remains after chemical doping (Figure 2a, right panel) has a valley polarization of about 25\% (Figure 2b, right panel). We consistently observe the emergence of strong valley polarization after anisole treatment in multiple samples (Supplementary Section S2). 

Next, we demonstrate the substrate independence of the effect of our treatment by repeating the measurements on an yttrium iron garnet (YIG) substrate. YIG is a magnetic insulator that was shown to effectively negatively dope $\text{MoS}_{2}$ monolayers at low temperatures, possibly due to dangling oxygen bonds at the YIG surface\cite{Peng2017}. As such, the total level of doping could be larger for monolayers on YIG due to additional doping from the substrate.

We exfoliated monolayers $\text{WS}_{2}$ onto polydimethylsiloxane (PDMS) stamps and deposited them onto the YIG substrates\cite{Castellanos-Gomez2014}. As before, the emission of the monolayers is strongly quenched after chemical doping and a valley polarization of about 20\%-40\% emerges (Figure 3, Supplementary Section S2). Compared to the monolayers on Si/SiO$_2$ substrates we conclude that these data do not indicate significant additional doping from the YIG substrate.

To assess the spatial homogeneity of the doping, we characterize the photoluminescence and valley polarization of a relatively large-area monolayer flake on YIG (Figure 3a-b). In most parts of the flake, we observe a valley polarization of about 40\%. In addition, at multiple spots in the monolayer, we observe an enhanced photoluminescence and reduced valley polarization. A comparison with an atomic force microscope topography image (Figure 3c) shows that these spots are associated with wrinkles in the flake. Spectrally, the spots are characterized by the simultaneous presence of an exciton resonance and a trion resonance, with the exciton resonance rapidly vanishing as we move off the spot (Figure 3d, Supplementary section S3) and the trion brightness and polarization remaining approximately constant (Figure 3e-f). The stronger exciton emission at wrinkles indicates that the doping is less effective, possibly resulting from the restricted physical access to the monolayer at wrinkles or from a decreased substrate-induced doping due to the increased substrate-monolayer distance. In addition, the exciton and trion formation could be altered at the wrinkles as a result of local strain \cite{Harats2020}.   

Strikingly, the excitons at the wrinkles also attained a strong valley polarization, as can be seen from the spectra in Figure 3d. We extend an existing rate equation model\cite{Lien2019} to argue that this is the result of the doping-induced conversion of excitons into trions (Figure 4a). This conversion acts as a decay channel for the excitons, enhancing their valley polarization and quenching their photoluminescence. The model predicts that the excitonic valley polarization starts to increase strongly when the conversion rate into trions $\Gamma_{\text{T}\leftarrow \text{X}}$ becomes comparable to the intervalley scattering rate $\Gamma_{\text{iv,X}}$ (Figure 4b, green line). Since $\Gamma_{\text{T}\leftarrow \text{X}}$ is proportional to the electron density as described by a law of mass-action\cite{Ross2013,Siviniant1999}, indeed an emergent exciton polarization is expected when doping is strong.

Strongly valley-polarized excitons are expected in the limit of large doping (Figure 4b). For our flakes, doping is strongest in the flat areas away from the wrinkles as reflected by the low photoluminescence in these areas. Because we are unable to spectrally distinguish the weak exciton emission from the dominant trion emission in these areas, we analyse the valley polarization of the integrated photoluminescence spectrum using our APD. When plotting the local valley polarization against the local photoluminescence (Figure 4c), we observe a non-monotonous behaviour with a maximum at low photoluminescence. According to our model, this maximum occurs because the exciton valley polarization (green line in Figure 4b) increases with doping while the exciton photoluminescence vanishes. As a result, the trion contribution (red line) starts to dominate the total signal (black line). These results highlight that the exciton valley polarization becomes large because of the rapid conversion into trions.  
  
On wrinkles, we observe that the excitons have a lower valley polarization than the trions (Figure 3d). In contrast, our model predicts that the local valley polarization of the trions cannot exceed that of the excitons even at low doping (Figure 4b, Supplementary Section S4). This indicates that the observed spectra on wrinkles are a result of spatial averaging over less-doped, wrinkled areas with a strong exciton contribution and strongly-doped surrounding areas with a dominant trion emission (Supplementary Section S5). Such averaging is expected from the diffraction-limited optical spotsize of our confocal microscope (diameter: $\sim500$ nm). 
  
In summary, we have demonstrated that chemical doping with anisole is an effective method to generate highly valley-polarized excitons and trions in monolayer $\text{WS}_{2}$ at room temperature. The emission spectrum of as-prepared monolayers is characterized by a bright exciton resonance that exhibits no valley polarization. After chemical doping, a trion resonance appears with a polarization up to 40\%. The doping is less efficient at wrinkled areas, which are marked by the simultaneous presence of exciton and trion resonances. The excitons have a robust valley polarization, which we attribute to the  rapid conversion into trions induced by the doping. A rate equation model captures the quenching-induced valley polarization, indicating the presence of excitons with a higher polarization than trions in the limit of maximal quenching. Our results shed light on the effect of the doping-controlled conversion between excitons and trions on the valley polarization in single layers of $\textrm{WS}_{2}$ and highlight that valley polarization by itself does not necessarily reflect optovalleytronic potential, since a strongly-quenched carrier lifetime and emission may constrain its application in devices. 

\section*{Methods}

\subsection*{Experimental setup.} Our samples are excited by a lowpass-filtered 594 nm OBIS laser (Coherent) of which we control the polarization using achromatic half- and quarter-wave plates (Thorlabs). A 50X, NA=0.95 (Olympus) objective focuses the laser to a diffraction-limited spot and collects the emission from the sample. The handedness of the excitation and detection is controlled by a second quarter-wave plate, which projects both circular polarizations of the photoluminescence onto two orthogonal linear polarizations of which we select one with the polarizer. The emission is longpass filtered (2x Semrock, BLP01-594R-25) to eliminate the laser reflection. We use a mirror on a computer-controlled flipmount to switch between a fiber-coupled spectrometer (Kymera 193 spectrograph with a cooled iVac 324 CCD detector) and an avalanche photodiode (APD, Laser Components) for the detection of the photoluminescence. Before the emission is detected by the APD, it is filtered with a pinhole and bandpass filter (Semrock, FF01-623/32-25). The sample is mounted on an xyz-piezo stage (Mad City Labs, Nano-3D200FT) to allow nanoscale positioning of the sample. An ADwin Gold II was used to control the piezo stage and read out the APD. All measurements were performed at room temperature.\\
\subsection*{Sample fabrication.} The $\textrm{WS}_{2}$ monolayers were exfoliated from commercially-purchased bulk crystals (HQ Graphene) on PDMS stamps, and were transferred to Si/$\textrm{SiO}_{2}$ and YIG chips. The 245 nm thick YIG films were grown on a gadolinium gallium garnet (GGG) substrate via liquid phase epitaxy and were purchased at Matesy gmbh. YIG samples were sonicated in acetone and cleaned in IPA before stamping.

\bibliography{Carmiggelt2020_library}

\section*{Acknowledgements}

This work was supported by the Netherlands Organisation for Scientific Research (NWO/OCW), as part of the Frontiers of Nanoscience program, through the Startup grant 740.018.012, and by the Kavli Institute of Nanoscience Delft. The authors thank N. Papadopoulos, M.\ Lee for their advice on fabrication and I.\ Komen, L.\ Kuipers for experimental help and discussions.

\section*{Author contributions statement}
J.C. and T.S. conceived the experiment. M.B. fabricated the samples. J.C. and M.B. conducted the experiment and analysed the results. M.B. created the figures. J.C. and T.S. wrote the manuscript. All authors reviewed the manuscript. T.S. supervised the project.

\section*{Additional information}
\textbf{Data availability}: The numerical data plotted in the figures in this work are available in Zenodo\cite{Carmiggelt2020}.
\textbf{Supplementary information}: available at http://www.nature.com/scientificreports
\textbf{Competing interests}: The authors declare no competing financial interests.


\begin{figure*}[ht]
	\includegraphics[]{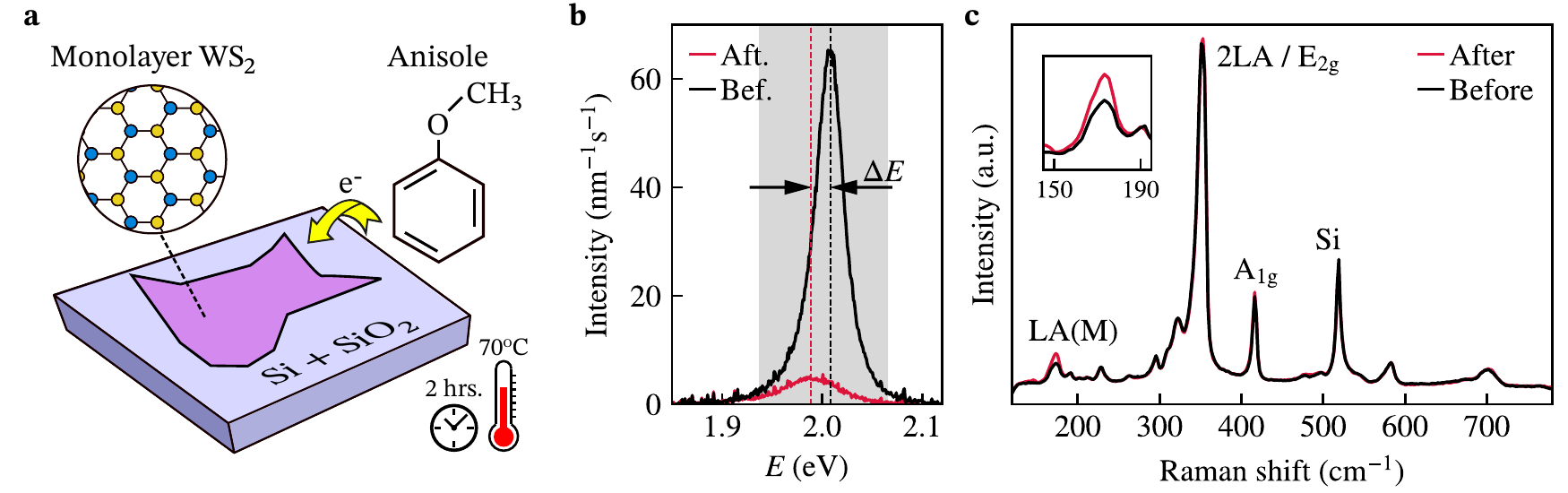}
	\caption{Controlling the photoluminescence properties of monolayer $\text{WS}_{2}$ via chemical doping. (a) $\text{WS}_{2}$ monolayers on Si/SiO$_2$ substrates become n-doped by treating them with anisole for 2 hours at 70 $^{\circ}$C. The insets show the chemical structures of  $\text{WS}_{2}$ and anisole. (b) Photoluminescence spectrum of a monolayer $\text{WS}_{2}$ before and after the anisole treatment. The treatment quenches the neutral exciton resonance, leading to the emergence of the trion resonance. The spectrum before (after) treatment was taken at 4 $\upmu$W (40 $\upmu$W) off-resonant laser excitation ($E=2.331$ eV, $\lambda=$ 532 nm). The shaded area indicates the transmission window of the bandpass filter used for the maps in Figure 2. (c) Raman spectra before and after the treatment of the same monolayer as in (b), at 514 nm laser excitation. The inset shows the enhanced intensity of the longitudinal acoustic LA(M) phonon mode, attributed to the adsorption of the anisole molecules. Both spectra are averages over multiple positions of the flake, which all show the same mode enhancement.}
\end{figure*}

\begin{figure}[ht]
	\includegraphics[]{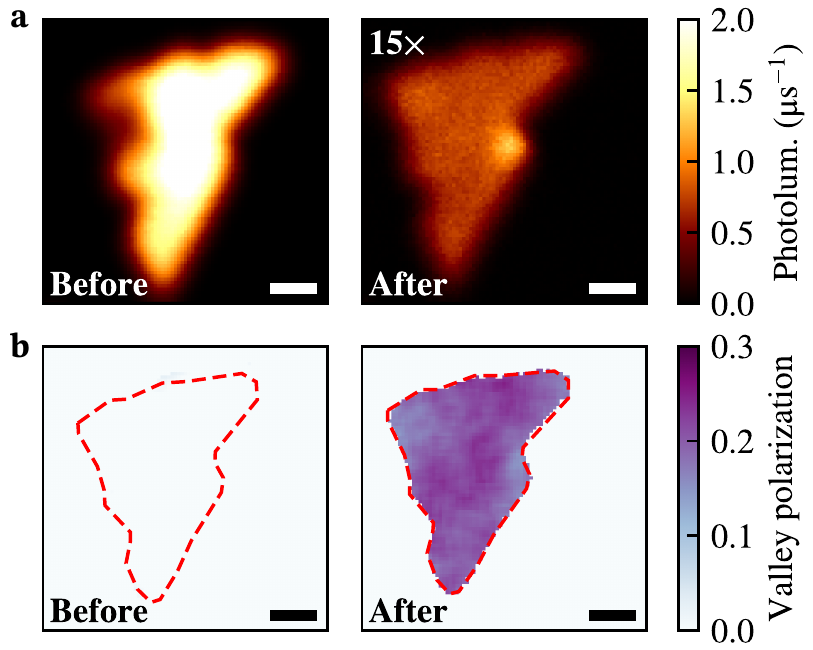}
	\caption{Spatial maps of the photoluminescence (a) and valley polarization (b) of a monolayer $\text{WS}_{2}$ before and after chemical doping with anisole. The treatment quenches the brightness of the flake and gives rise to strongly valley-polarized emission. The flake was exfoliated on a Si/SiO$_2$ substrate and excited near-resonance ($E=2.087$ eV, $\lambda=$ 594 nm, 4 $\upmu$W). Scale bar: 2 $\upmu$m.}
\end{figure}

\begin{figure*}[ht]
	\includegraphics[]{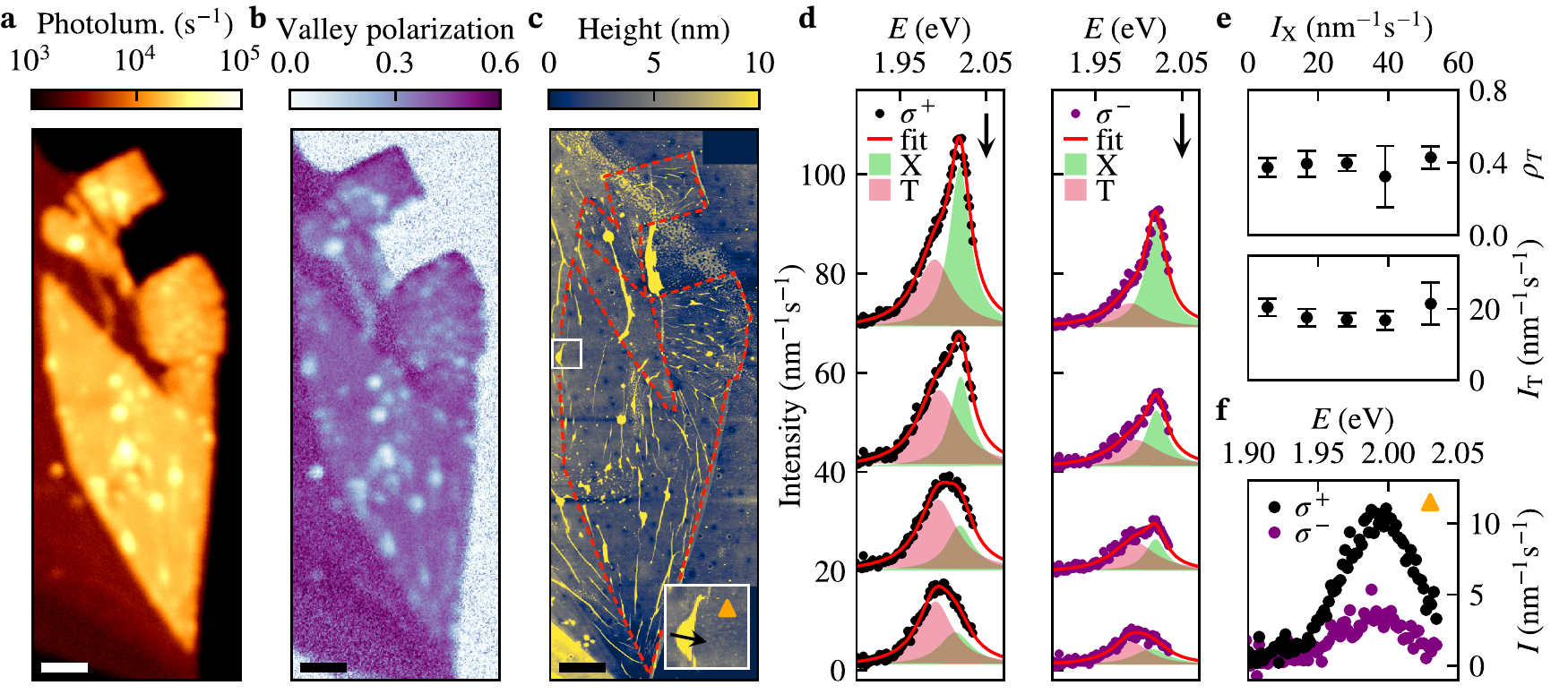}
	\caption{Spatial characterization of the exciton and trion emission of a chemically-doped monolayer $\text{WS}_{2}$ on an yttrium iron garnet (YIG) substrate. (a-b) Spatial maps of the photoluminescence and valley polarization under near-resonant excitation (594 nm, 200 $\upmu$W) after chemical doping. The sample was submerged in liquid anisole for 12 hours at room temperature and vacuum-annealed for 6 hours (400 $^{\circ}$C, $<$1 mTorr) to remove contaminants. Multilayer areas of the flake surrounding the monolayer are identified by their low brightness due to their indirect bandgap \cite{Splendiani2010} and large polarization\cite{Su2017}. A comparison with the atomic force microscope image in (c) shows that spots with increased photoluminescence and reduced valley polarization occur at wrinkles of the monolayer. (d) Emission spectra at different locations close to a wrinkle indicated by the black arrow in the inset of (c). Lorentzian fits of the trion (red) and exciton (green) resonances reveal the simultaneous presence of trion and exciton emission at wrinkles. (e) Average trion brightness and valley polarization plotted against the local exciton photoluminescence at different wrinkles. (f) Typical $\sigma_+$ and $\sigma_-$ emission spectra of trions in flat parts of the flake, obtained at the location indicated by the triangle in the inset of (c), corresponding to a valley polarization of about 40\%. Scale bar: 5 $\upmu$m.}
\end{figure*}

\begin{figure}[ht]
	\includegraphics[]{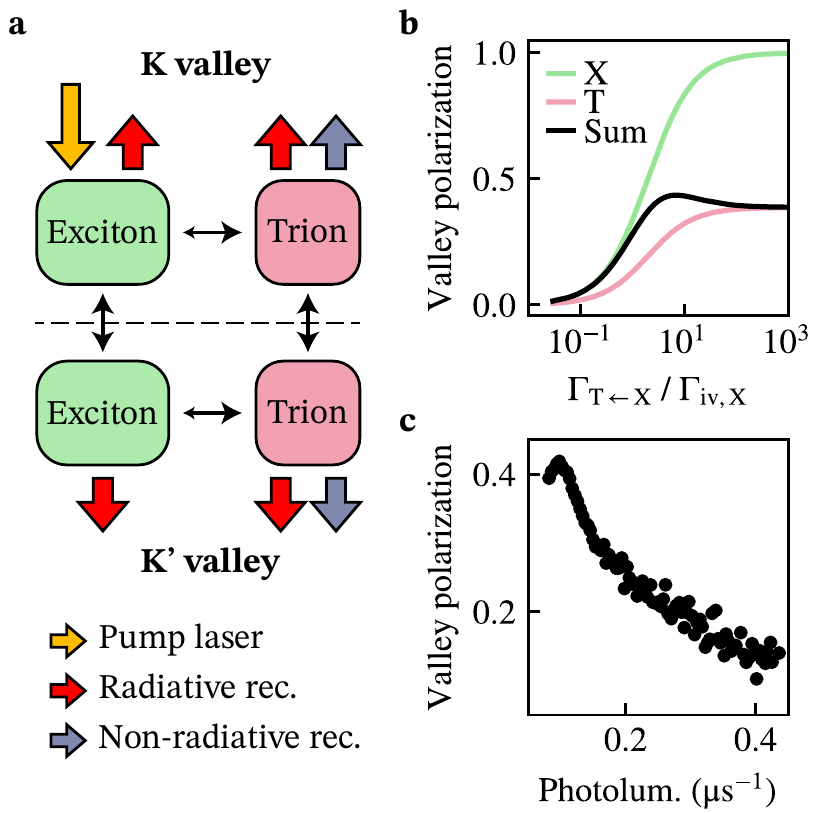}
	\caption{Doping-controlled valley polarization of excitons and trions. (a) Schematic depiction of the rate equation model used to describe the optically detected valley polarization. Excitons are created by valley-selective optical excitation, after which they can decay radiatively, scatter between the valleys at a rate $\Gamma_\text{iv,X}$, or change into trions at a doping-controlled rate $\Gamma_{\text{T}\leftarrow \text{X}}$. The trions can scatter between the valleys, decay radiatively or non-radiatively, and change back into excitons. (b) Valley polarization of excitons, trions, and their photoluminescence-weighted average as a function of $\Gamma_{\text{T}\leftarrow \text{X}}/\Gamma_\text{iv,X}$ calculated using the rate equation model shown in (a). (c) Valley polarization versus photoluminescence extracted by averaging data from individual pixels in the monolayer area of Figure 3a-b.}
\end{figure}

\end{document}